%% file: UO2_oxygen_defects.tex
\documentclass[aps,prl,twocolumn,superscriptaddress,groupedaddress]{revtex4}

\usepackage{graphicx}  
\usepackage{wrapfig}
\usepackage{dcolumn}   
\usepackage{bm}        
\usepackage{amssymb}   
\usepackage{multirow}
\usepackage{anyfontsize}
\begin{document}

\hspace{5.2in} \mbox{}

\title{Temperature-induced atomic-structure modifications in UO$_{2}$ and UO$_{2+x}$}
\input author.tex

\date{\today}

\begin{abstract}
UO$_{2}$ and UO$_{2.07}$ were characterized from 25--1000$\,^{\circ}\mathrm{C}$ using neutron total scattering in order to evaluate effects of temperature and phase boundaries on local and average structures. Analyses of unit cell parameters showed that both materials exhibit very similar thermal expansion behavior and thermal expansion data lay along the upper bound of uncertainty for standard empirical models, indicating a slightly faster thermal expansion rate. Atomic displacement parameters of UO$_{2.07}$ showed evidence for U$_{4}$O$_{9}$ phase boundaries in accordance with the established phase diagram, despite the suppression of U$_{4}$O$_{9}$ superlattice peaks in the measured diffraction patterns. Pair distribution functions revealed that the differences in local structure between UO$_{2}$ and UO$_{2.07}$ were very small and PDF features are dominated by thermal effects. The rate of contraction of the first nearest-neighbor U-O distances of UO$_{2}$ and UO$_{2.07}$ between 25--1000$\,^{\circ}\mathrm{C}$ were shown to agree with molecular dynamics simulations and local structure analyses previously performed on UO$_{2}$ above 1000$\,^{\circ}\mathrm{C}$.
\end{abstract}

\maketitle

\section{1. Introduction}
UO$_{2}$, as a nuclear fuel, incorporates extraordinary amounts of atomic disorder during and after in-pile operation. Oxygen defects, specifically interstitials, play a key role in the structural evolution of this material, especially at high temperature. Oxygen interstitials in oxidized UO$_{2}$ are known to cluster and can ultimately influence phase stability and important physiochemical properties, such as diffusivity \cite{n1}. Therefore, accurate assessment of temperature-induced structural changes is necessary in order to better predict the behavior of engineering properties under off-normal conditions.

The structure of UO$_{2}$ has been studied for decades; however, there is increasing evidence to show that stoichiometric UO$_{2}$ exhibits structural subtleties that may account for unusual and unexpected properties, such as anisotropic thermal conductivity \cite{n2}. Pure, stoichiometric UO$_{2}$ exhibits the fluorite structure (space group $Fm\overline{3}m$) at room temperature with uranium and oxygen occupying 4\textit{a} and 8\textit{c} sites, respectively. This structure has been shown to persist to temperatures as high as $\sim$2865$\,^{\circ}\mathrm{C}$, where melting occurs. 

Despite this, recent analyses of UO$_{2}$ at 1000$\,^{\circ}\mathrm{C}$ show that the local structure is inconsistent with ideal fluorite structure symmetry and the material may exhibit local structural distortions induced by anion sublattice re-structuring \cite{n3}. Studies further suggest that the rate of thermal expansion for stoichiometric UO$_{2}$ is higher than previously assumed \cite{n4}. These investigations highlight the need to revisit and re-evaluate the structural properties of this important energy material.

This study investigates how temperature and moderate hyper-stoichiometry influence atomic arrangements in fluorite-structured uranium oxide. Emphasis is placed on correlating local-structure modifications to average-structure modifications as they relate to temperature variation and phase changes. Changes in short-range atomic structures are characterized through inspection of atomic displacement parameters (ADPs) from diffraction, and pair distribution function (PDF) analysis of neutron total scattering data.

The structures of UO$_{2}$ and UO$_{2.07}$ were investigated from room temperature to 1000$\,^{\circ}\mathrm{C}$. This is unique from recent PDF studies of UO$_{2}$ that focus exclusively on higher temperatures ($\ge$1000$\,^{\circ}\mathrm{C}$) \cite{n3,n5}. Results from this study show that phase transitions can be identified by statistically-significant fluctuations in ADP behavior. PDF analyses reveal that local structural modifications induced by both temperature and oxidation are extremely subtle and require detailed structural modeling in order to reveal oxygen defect ordering schemes in hyper-stoichiometric UO$_{2}$ (UO$_{2+x}$).

\section{2. Experimental}
\subsection{2.1. Sample Preparation}
Polycrystalline powders were derived from dense pellets made with UO$_{2+x}$ ($\sim$UO$_{2.16}$) feedstock material purchased from International Bio-analytical Industries Inc. USA. Dense microcrystalline pellets were prepared by spark plasma sintering (SPS) at 1300$\,^{\circ}\mathrm{C}$ for either 5 minutes or 30 minutes under a load of 40 MPa using graphite dies. The use of graphite dies ensured that the uranium oxide pellets were reduced \textit{in situ} during the sintering process, with the degree of reduction being dependent on the sintering time. The as-prepared pellets were characterized by X-ray diffraction (XRD) and stored in inert gas atmosphere. Additional details regarding the pellet preparation process and XRD measurements are provided elsewhere \cite{n6}. The powder neutron total scattering samples were prepared by grinding the dense pellets in inert gas atmosphere. Samples referred to as \textit{UO$_{2}$} and \textit{UO$_{2.07}$} were derived from pellets sintered for 30 minutes and 5 minutes, respectively.

\subsection{2.2. Thermogravimetric Analysis}
Weight gain data were collected through thermogravimetric analysis (TGA). TGA was performed within the Notre Dame Materials Characterization Facility (MCF) using a Setaram LABSYS evo TGA-DSC instrument following a procedure modeled after ASTM C-1453-00 \cite{n7}. Approximately 10-50 mg quantities of powder were loaded into 100 $\mu$L alumina crucibles and heated to 900$\,^{\circ}\mathrm{C}$ under flowing gas atmosphere. Samples were run under synthetic air with a constant flow rate of 40 mL/min. All loaded samples were equilibrated at room temperature prior to tarring and increasing temperature. The heating rate was set to 10$\,^{\circ}\mathrm{C}$/min and the maximum temperature, 900$\,^{\circ}\mathrm{C}$, was held for three hours to allow for complete oxidation of samples to U$_{3}$O$_{8}$. Oxidation was determined to be complete when weight gain saturated and when a large, exothermic peak was observed in the differential scanning calorimetry (DSC) curve. The large exothermic peak denoted completion of the transformation to U$_{3}$O$_{8}$. Weight precision of the balance was $\sim$0.01\%.

\subsection{2.3. Inductively Coupled Plasma Mass Spectrometry (ICP-MS)}
All sample digestions and ICP-MS analyses were performed in a Midwest Isotope and Trace Element Research Analytical Center (MITERAC) class-1000 clean room at the University of Notre Dame. Digestions were conducted using oxidized U$_{3}$O$_{8}$ powders from TGA. Approximately 10--20 mg quantities of powder were loaded into 15 mL Savillex Teflon vials along with 3 mL of concentrated nitric acid and $\sim$0.5 mL of double-distilled, concentrated hydrofluoric acid. The mixtures were heated at 150$\,^{\circ}\mathrm{C}$ until samples were completely dissolved and all liquid had evaporated. Dried and concentrated samples were re-dissolved in 0.5 mL mixtures of 8-molar (8M) nitric acid and 0.1M hydrofluoric acid.

These samples were subjected to uranium ion exchange in order to selectively separate uranium from the impurity elements of interest that exist in low concentrations. The uranium separation procedure was modeled after experiments described in reference \cite{n8}. Uranium separations were performed using UTEVA resins that were pre-conditioned with milli-Q water, 3M nitric acid, and finally, mixtures of 8M nitric acid and 0.1M hydrofluoric acid. Samples obtained after ion exchange were dried at 110$\,^{\circ}\mathrm{C}$ in Teflon vials, rinsed, and re-suspended in 5 mL of 2\% nitric acid spiked with arsenic internal standard for ICP-MS measurements.

Digested samples and diluted standards were measured with a high-resolution Attomm ICP-MS instrument by Nu Instruments. Reported concentrations represent averaged values from eleven individual measurements. Results for UO$_{2}$ and UO$_{2.07}$ samples were combined because both materials were synthesized using the same feedstock material and concentrations were expected to be low. Detection limits were estimated with linear calibration curves for high-purity elemental standards in the range 0.005--175 $\mu$g/L using linear regression coefficients in excess of 0.99. 

\subsection{2.4. Neutron Total Scattering}
Neutron total scattering measurements were performed at the Nanoscale Ordered Materials Diffractometer (NOMAD) beamline at the Spallation Neutron Source (SNS) at Oak Ridge National Laboratory. Powder UO$_{2}$ and UO$_{2.07}$ samples ($\sim$1 g each) were loaded in a glovebox filled with inert gas in order to avoid exposure of the powder samples to air. Each sample was double-encapsulated by loading into a quartz nuclear magnetic resonance (NMR) tube that was subsequently loaded into a vanadium can. Loaded vanadium cans were sealed inside of the glovebox prior to removal for measurements at the beamline.

Each sample was measured at room temperature and at elevated temperature inside of an Institut Laue-Langevin (ILL)-type vacuum furnace. High vacuum (10$^{-6}$--10$^{-7}$ Torr) was maintained throughout the experiments. Data were collected for 30 minutes at each temperature indicated in Figure 1. Shorter measurement times of 2 minutes were also used to collect data \textit{in situ} as the samples were heated. 30-minute collection times yielded higher-quality data that enabled pair distribution function (PDF) analysis. Lower-quality data from 2-minute measurements enabled diffraction analysis because of the strong Bragg scattering signal, but counting statistics for these 2-minute data were not high enough to permit accurate PDF analysis.

\begin{figure}[!htb]
	\begin{center}
		\includegraphics[width=0.48\textwidth,trim=2 2 2 2,clip]{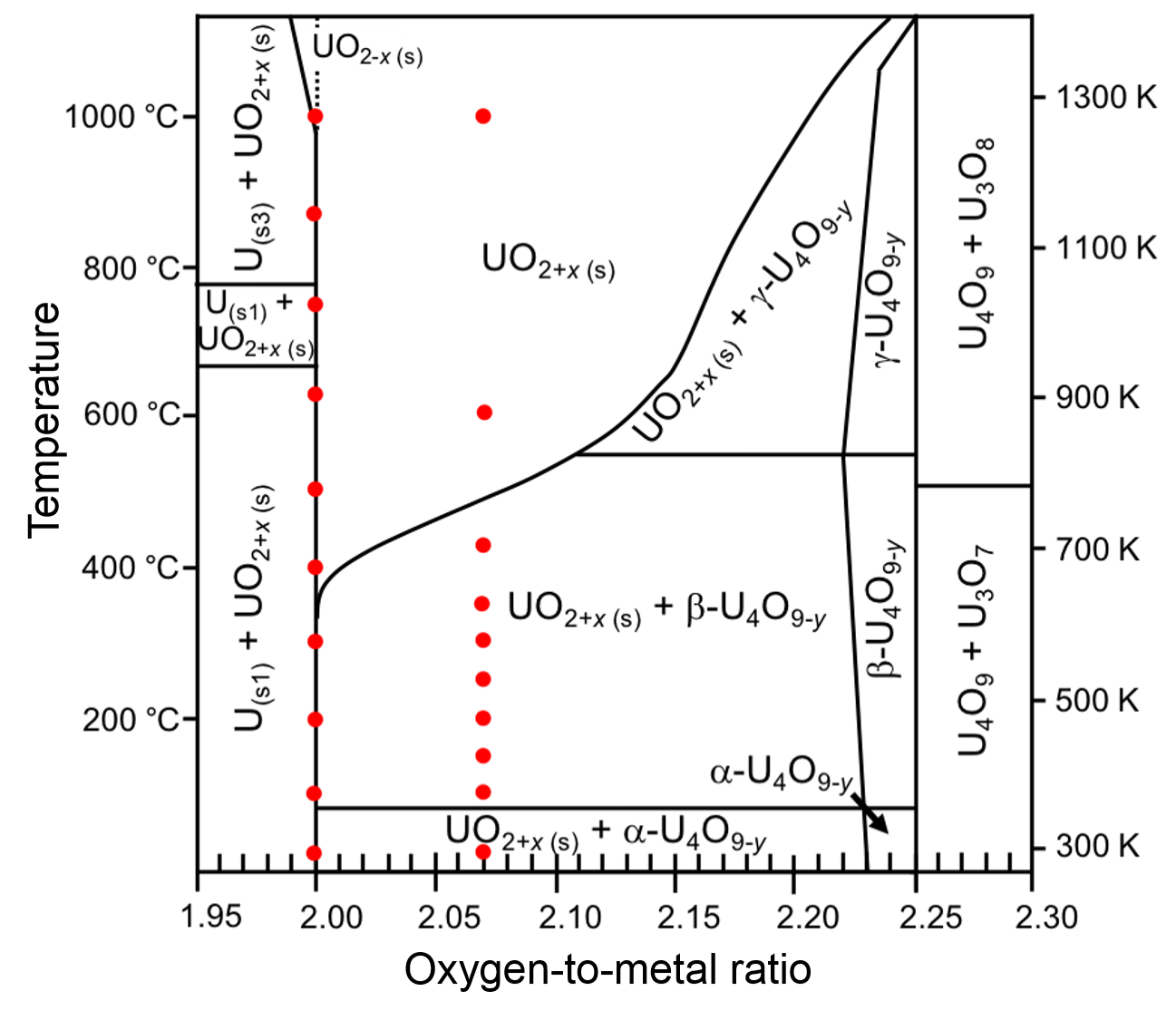}
		\caption{\label{Figure:1} Phase diagram of uranium oxide adapted from reference \cite{n14}. Red points denote temperatures at which neutron total scattering data were collected for UO$_{2}$ and UO$_{2.07}$ samples.}
	\end{center}
\end{figure}

NOMAD detectors were calibrated using diamond powder and data were normalized using measured scattering intensity from a solid vanadium rod. Total scattering measurements were corrected for multiple-scattering and absorption effects and data were normalized to absolute scale by taking into account densities and packing fractions of the measured samples. Corrected total-scattering structure factors, $F(Q)$, were converted into differential correlation functions, $D(r)$, using the relation:
\begin{displaymath}
D(r) = 4 \pi r \rho_{0}\,G(r)
\end{displaymath}
where $G(r)$ is the total radial distribution function defined as:
\begin{displaymath}
G(r)=\frac{1}{(2\pi)^3\rho_{0}} \int_{Q_{min}}^{Q_{max}} 4 \pi Q^{2}\,F(Q)\,\frac{sin(Qr)}{Qr}\,dQ
\end{displaymath}
where $\rho_{0}$ is the average number density of the material, $Q$ is the scattering vector of length $4\pi sin(\theta)/\lambda$ for a neutron of wavelength $\lambda$ scattered at an angle $2\theta$, and $r$ is real-space distance. $Q_{min}$ and $Q_{max}$ were set to 0.3 and 31.4 \AA$^{-1}$, respectively.

\subsection{2.5. Data Analysis}
Neutron total scattering data were analyzed by both Rietveld refinement and PDF analysis. Rietveld refinement of diffraction patterns was performed using the GSAS software \cite{n9}. Instrument parameters for GSAS were obtained by refining the diffraction patterns of a National Institute of Standards and Technology (NIST) silicon powder standard. Diffraction patterns of both UO$_{2}$ and UO$_{2.07}$ were fit with the fluorite structure (space group $Fm\overline{3}m$) with uranium located at the 4\textit{a} site and oxygen at the 8\textit{c} site. In the case of UO$_{2.07}$, oxygen interstitials ($x$ = 0.07) were added to the 4\textit{b} octahedral interstitial site. A total of 12 parameters were refined. These included a 6-coefficient background polynomial, scale factor, zero-point, isometric unit cell parameter, and isotropic atomic displacement parameters (ADPs) for 4\textit{a}, 8\textit{c}, and 4\textit{b} sites.

PDFs in the form of the $D(r)$ functions were modeled by small-box refinement performed with the PDFgui software \cite{n10}. Fitted models were identical to the models used for Rietveld refinement. Small-box refinements were performed with a total of 6 refineable parameters. These included a scale factor, isometric unit cell parameter, correlated motion parameter, and isotropic ADPs for 4\textit{a}, 8\textit{c}, and 4\textit{b} sites. The correlated motion parameter is unique to small-box refinement and reproduces effects of correlated, short-range atomic motion. Overall material compositions in Rietveld and small-box refinements were fixed to experimentally-determined stoichiometry values (see Section 3.1).

\section{3. Results and Discussion}
\subsection{3.1. Stoichiometry Determination}
Stoichiometry measurements were performed by two methods. The first method is based on unit cell parameter measurements. The UO$_{2}$ matrix shrinks upon oxidation owing to electronic structure modifications, such as the incorporation of U$^{5+}$ ions, which exhibit smaller ionic radii compared to U$^{4+}$ ions. The degree of lattice contraction in the low oxygen-to-metal regime ($x$ $<$ 0.12 for UO$_{2+x}$) can be directly related to the amount of oxidation through the use of empirical equations. The unit cell parameter, \textit{a}, was directly related to the oxygen-to-metal ratio (O:M) by the expression \cite{n11}:
$$a = (5.4705-0.1306 x) \AA $$
where $x$ is the degree of deviation from perfect stoichiometry for UO$_{2+x}$ (\textit{i.e.}, $x$ = 0 for stoichiometric UO$_{2}$). Applying unit cell parameters measured by laboratory X-ray diffraction of the dense pellets yielded O:M ratio values of 1.996(4) and 2.073(3) for the UO$_{2}$ and UO$_{2.07}$ samples, respectively. Unit cell parameters were also derived from neutron total scattering measurements. Unit cell parameters measured by neutron diffraction during the total scattering experiments yielded O:M ratio values of 1.995(4) and 2.068(1) for the UO$_{2}$ and UO$_{2.07}$ samples, respectively.

The second method used to calculate stoichiometry was applied after the neutron total scattering measurements using the ICP-MS and TGA data following a procedure modeled after ASTM C-1453-00 \cite{n7}. This method is based on the assumption that weight gain during oxidation of UO$_{2}$ to U$_{3}$O$_{8}$ is solely from the ingress of oxygen into the material. It is therefore important to quantify impurities in order to differentiate weight gain of UO$_{2}$ oxidation from that of oxidation of impurity elements. Twenty non-volatile impurity elements were identified and quantified by ICP-MS. Quantification of impurity elements was limited to twenty elements that were most likely to be present in UO$_{2}$ based on reported findings for similar sample types (see reference \cite{n12}). Measured impurity concentrations are shown in Table 1. Values are reported in units of $\mu$g of impurity element per gram of U$_{3}$O$_{8}$ from ignition (TGA). Values represent averages from eleven measurements performed on distinct ICP-MS samples.

\begin{table}[]
	\begin{tabular}{cccc}
		\multicolumn{1}{p{1.6cm}}{$\,\,\,\,$element} & \multicolumn{1}{p{1.6cm}}{$\mu$g element per g U$_{3}$O$_{8}$} & \multicolumn{1}{p{2.2cm}}{$\,\,\,\,\,\,\,\,\,\,$g oxide per {\tiny .}$\,\,\,\,\,\,\,\,\,\,\,$g element} & \multicolumn{1}{p{1.6cm}}{g oxide per {\tiny .}$\,\,\,$g U$_{3}$O$_{8}$} \\ \hline
		\multicolumn{1}{c}{Li} & \multicolumn{1}{c}{1.28} & 2.15 & 2.75E-06 \\
		\multicolumn{1}{c}{Mg} & \multicolumn{1}{c}{17.46} & 1.66 & 2.90E-05 \\
		\multicolumn{1}{c}{Al} & \multicolumn{1}{c}{10.51} & 1.89 & 1.99E-05 \\
		\multicolumn{1}{c}{Cr} & \multicolumn{1}{c}{2} & 1.46 & 2.92E-06 \\
		\multicolumn{1}{c}{Mn} & \multicolumn{1}{c}{1.55} & 1.58 & 2.45E-06 \\ 
		\multicolumn{1}{c}{Fe} & \multicolumn{1}{c}{69.07} & 1.43 & 9.88E-05 \\ 
		\multicolumn{1}{c}{Ni} & \multicolumn{1}{c}{10.65} & 1.27 & 1.35E-05 \\ 
		\multicolumn{1}{c}{Co} & \multicolumn{1}{c}{0.1} & 1.41 & 1.41E-07 \\ 
		\multicolumn{1}{c}{Cu} & \multicolumn{1}{c}{5.72} & 1.25 & 7.15E-06 \\ 
		\multicolumn{1}{c}{Zn} & \multicolumn{1}{c}{11.46} & 1.24 & 1.42E-05 \\ 
		\multicolumn{1}{c}{Zr} & \multicolumn{1}{c}{3.45} & 1.35 & 4.66E-06 \\
		\multicolumn{1}{c}{Mo} & \multicolumn{1}{c}{1.91} & 1.5 & 2.87E-06 \\ 
		\multicolumn{1}{c}{Cd} & \multicolumn{1}{c}{0.11} & 1.14 & 1.25E-07 \\
		\multicolumn{1}{c}{In} & \multicolumn{1}{c}{0.05} & 1.21 & 6.10E-08 \\ 
		\multicolumn{1}{c}{Sn} & \multicolumn{1}{c}{38.48} & 1.27 & 4.89E-05 \\
		\multicolumn{1}{c}{Ba} & \multicolumn{1}{c}{0.8} & 1.12 & 8.96E-07 \\
		\multicolumn{1}{c}{La} & \multicolumn{1}{c}{0.05} & 1.17 & 5.90E-08 \\
		\multicolumn{1}{c}{Ce} & \multicolumn{1}{c}{0.03} & 1.17 & 3.50E-08 \\
		\multicolumn{1}{c}{Gd} & \multicolumn{1}{c}{3.1} & 1.15 & 3.57E-06 \\
		\multicolumn{1}{c}{Pb} & \multicolumn{1}{c}{1.3} & 1.15 & 1.50E-06 \\ \hline
		&  & sum: & 2.53E-04 \\
		&  & standard deviation: & 2.37E-05 \\
	\end{tabular}
	\caption{\label{Table 1:} Impurity concentrations in UO$_{2}$ and UO$_{2.07}$ samples as measured by inductively coupled plasma mass spectrometry (ICP-MS). Concentrations represent the average of eleven measurements. Oxide conversion factors (g oxide/g element) were obtained from reference \cite{n7}.}
\end{table}

Weight gain values from TGA and impurity concentrations from ICP-MS were used to calculate uranium content, U, in weight percent with the expression \cite{n7}:
\begin{displaymath}U = \Bigg(\frac{0.8480 \times (w_{F}-w_{F} \times I_{N})}{w_I}\Bigg) \times 100 - I_{L} \end{displaymath}
where 0.8480 is a conversion factor, $w_{I}$ is sample weight prior to ignition, $w_{F}$ is sample weight after ignition, $I_{N}$ is the total of measured nonvolatile impurities in grams of oxide per gram of U$_{3}$O$_{8}$, and $I_{L}$ is the total of nonvolatile impurities less than the lower detection limit. The latter value was assumed to be 0.01\% based on the recommendation in reference \cite{n7}. Uranium content values were used to calculate O:M values by the equation:
\begin{displaymath}O:M = \frac{(100-U) \times 238.03}{15.9994 \times U} \end{displaymath}
where 238.03 and 15.9994 are the atomic weights of uranium and oxygen, respectively. Calculated O:M values for UO$_{2}$ and UO$_{2.07}$ samples were 1.998(16) and 2.075(16), respectively. Uncertainty values were derived by error propagation. A comparison of the calculated stoichiometry measurements from the two methods (lattice parameter and combustion analysis) shows good agreement among the different calculations and indicates O:M values of $\sim$2.00 and $\sim$2.07 for UO$_{2}$ and UO$_{2.07}$ samples, respectively.

\subsection{3.2. Neutron Diffraction – Average Structures}
Neutron diffraction patterns of UO$_{2}$ and UO$_{2.07}$ samples show that both materials are well-ordered and exhibit fluorite-type structures (space group $Fm\overline{3}m$). Figure 2 shows the various diffraction patterns collected at room temperature for both samples from the six different detector banks located at different scattering angles. Each detector bank highlights a unique Q-range and exhibits distinct resolution. For example, the high-angle detector bank, bank 6 (2$\theta$ = 150.1$\,^{\circ}$), yields the highest resolution of all detector banks, but probes the smallest Q-range. A comparison of UO$_{2}$ and UO$_{2.07}$ data shows that all diffraction patterns show strong, sharp Bragg peaks and little to no peaks that are indicative of diffuse scattering from defects and short-range ordering.

\begin{figure}[!htb]
	\begin{center}
		\includegraphics[width=0.48\textwidth,trim=2 2 2 2,clip]{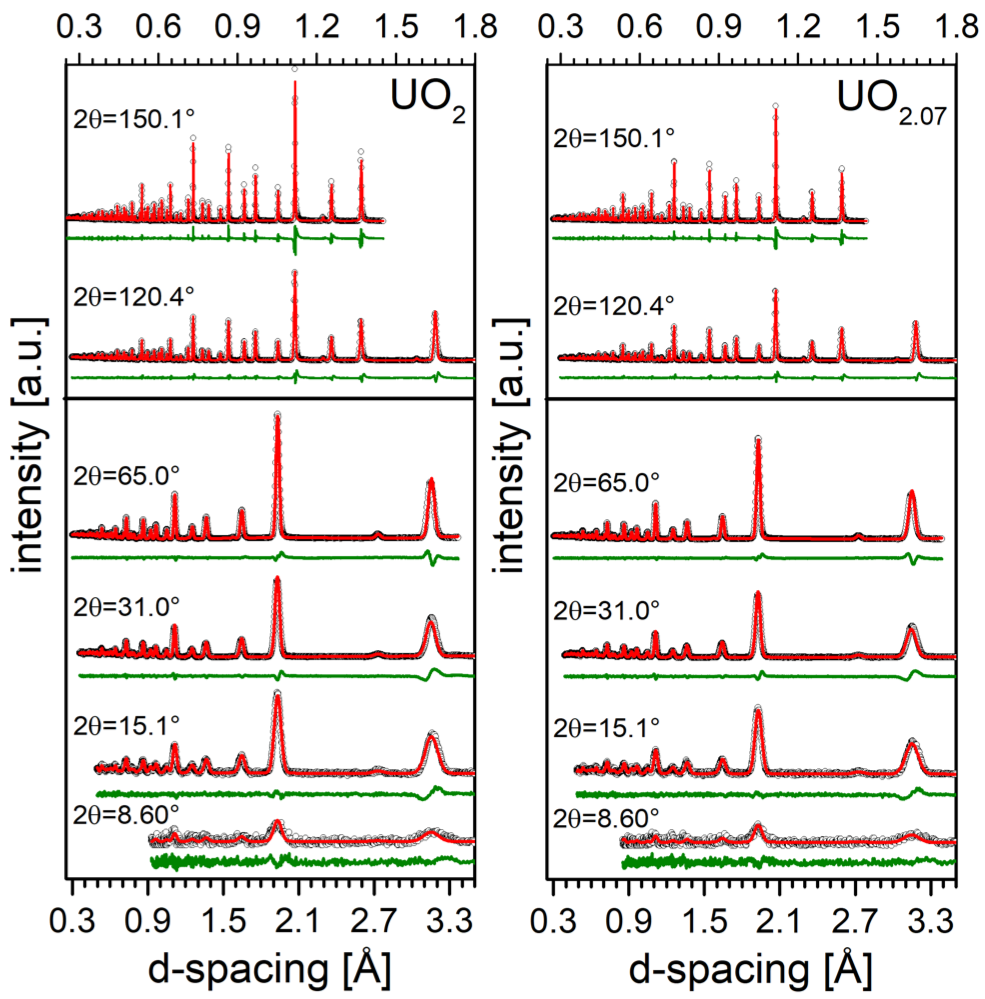}
		\caption{\label{Figure:2} Rietveld refinement fits to the various diffraction patterns of UO$_{2}$ and UO$_{2.07}$ collected at room temperature. Diffraction patterns from banks 1--4 (2$\theta$ = 8.60--65.0$\,^{\circ}$) are shown in the d-spacing range 0.3--3.3\AA\, whereas diffraction patterns from banks 5--6 (2$\theta$ = 120.4--150.1$\,^{\circ}$) are shown from 0.3--1.8\AA. Black circles are measured data, red lines are fitted fluorite structure models, and green lines represent the difference between data and the fitted models.}
	\end{center}
\end{figure}

Rietveld refinement of the diffraction patterns reveals that both materials are well represented with simple fluorite structure models at all temperatures. A fluorite structure model incorporating oxygen interstitials at the 4\textit{b} octahedral site (4\textit{b} site occupancy = 0.07) was fit to the diffraction patterns of UO$_{2.07}$ but yielded no improvement. Goodness-of-fit values, R$_{w}$, were identical (R$_{w}$ = 0.09) with and without incorporation of interstitials into the UO$_{2.07}$ structural model. Other defect and hyper-stoichiometric models, such as recently-derived U$_{4}$O$_{9}$-type structures \cite{n13}, were also attempted, but yielded no improvement in the fit. According to the established phase diagram \cite{n14}, UO$_{2.07}$ should exist as a two-phase UO$_{2+x (s)}$ + $\alpha$-U$_{4}$O$_{9-y}$ mixture below 50--100$\,^{\circ}\mathrm{C}$, but there is no evidence of superlattice U$_{4}$O$_{9}$ peaks in the UO$_{2.07}$ diffraction patterns at low temperature. U$_{4}$O$_{9}$-type models often yield diffraction peaks in the 2-2.5 $\AA$ d-spacing range, which are not present in the experimentally-measured patterns.

The Rietveld refinement fit results illustrate that U$_{4}$O$_{9}$-phase superlattice diffraction peaks are suppressed, despite the phase diagram indicating a UO$_{2+x (s)}$ + $\alpha$-U$_{4}$O$_{9-y}$ mixture. Similar observations were made previously \cite{n15} and are attributed to variations in sample synthesis routes. For example, it’s been demonstrated that rapid quenching can preserve high-temperature, single-phase UO$_{2}$+x (s) to room temperature \cite{n16}. Detailed analyses of atomic disorder parameters presented later suggest that quenched UO$_{2.07}$ contains hidden U$_{4}$O$_{9}$-type atomic ordering despite the apparent suppression of U$_{4}$O$_{9}$-type superlattice Bragg peaks.

Rietveld analyses were further used to extract unit cell parameters and measure atomic disorder indirectly through inspection of atomic displacement parameters (ADPs). Unit cell parameters measured with increasing temperature were converted into relative linear thermal expansion, $\mu$L/L0, in order to compare the data with the empirical thermal expansion model of Martin \cite{n17}, which was later expanded upon by Fink \cite{n18} (so-called Martin-Fink model). Relative linear thermal expansion values were referenced to values at 0$\,^{\circ}\mathrm{C}$ (273 K), \textit{i.e.}, $\Delta$L/L$_{0}$ = (L$_{T}$ – L$_{0}$)/L$_{0}$, where L$_{T}$ and L$_{0}$ are unit cell parameters at temperature T$\,^{\circ}\mathrm{C}$ and 0$\,^{\circ}\mathrm{C}$, respectively. Figure 3 shows the measured thermal expansion of UO$_{2}$ and UO$_{2.07}$ compared to the Martin-Fink prediction and recent results from Guthrie et al. \cite{n4}.

Measured thermal expansion data for both materials are in good agreement with the Martin-Fink model, although thermal expansion rates are closer to the upper bound of the Martin-Fink prediction. Guthrie et al. \cite{n4} recently reported a similar finding for stoichiometric UO$_{2}$ at more elevated temperatures ($\sim$1000--2500$\,^{\circ}\mathrm{C}$). The present data collected at lower temperatures (25--1000$\,^{\circ}\mathrm{C}$) are therefore complimentary to those results. Gurthrie et al. \cite{n4} attributed the faster rate of thermal expansion to improved experimental measurement resolution from synchrotron XRD, as the Martin-Fink model is based on data that are over 25 years old. Results from both Guthrie et al. \cite{n4} and the present study are in agreement with first-principles calculations that predict slightly higher thermal expansion coefficients \cite{n19}.

\begin{figure}[!htb]
	\begin{center}
		\includegraphics[width=0.48\textwidth]{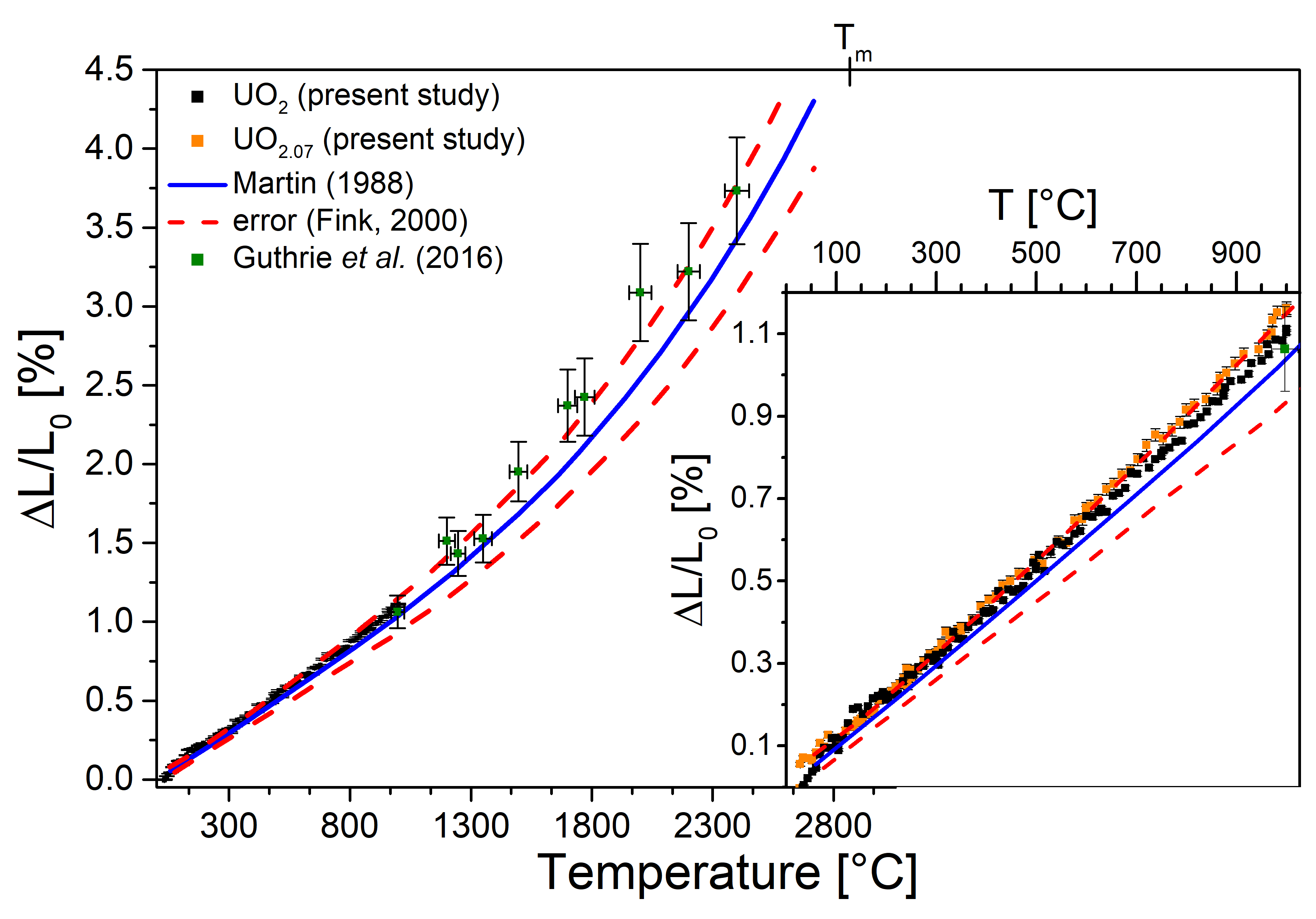}
		\caption{\label{Figure:3} Linear thermal expansion, $\Delta$L/L$_{0}$, of UO$_{2}$ and UO$_{2.07}$ from room temperature to the melting temperature (T$_{m}$). The inset shows an enlarged view of the region between 25--1000$\,^{\circ}\mathrm{C}$ to highlight data from the present study. Data are compared to the Martin-Fink model \cite{n17,n18} and synchrotron X-ray diffraction results from Guthrie et al. \cite{n4}.}
	\end{center}
\end{figure}

A comparison of thermal expansion data for UO$_{2}$ and UO$_{2.07}$ shows behavior in agreement with former assessments that hyper-stoichiometry has minimal effects on thermal expansion \cite{n17}. The datasets are identical within experimental uncertainty until $\sim$700$\,^{\circ}\mathrm{C}$ when the trends begin to increasingly deviate. At temperatures above $\sim$700$\,^{\circ}\mathrm{C}$, the rate of thermal expansion for UO$_{2.07}$ is slightly higher than that of UO$_{2}$. The data remain within limits of experimental uncertainty at the highest temperatures, but it is unknown if the two trends would deviate more at higher temperatures. Further studies of broader hyper-stoichiometric and temperature ranges are needed in order to better assess the influence of hyper-stoichiometry on thermal expansion rate.

Information regarding atomic disorder was derived from inspection of ADPs. Uranium and oxygen ADPs were modeled as isotropic (U$_{11}$=U$_{22}$=U$_{33}$; U$_{12}$=U$_{13}$=U$_{23}$) in accordance with fluorite structure symmetry constraints. ADPs, $U$, are related to instantaneous atomic displacement, $u$, by the expression: $U = <u^{2}>$. ADPs obtained from structural modeling are therefore a measure of the amount of atomic disorder at a particular Wyckoff site. Disorder can be caused by dynamic thermal atomic displacement and/or static atomic displacement from short-range ordering or defects. Figure 4 shows isotropic ADPs values of UO$_{2}$ compared to: (a) stoichiometric UO$_{2}$ results from Willis \cite{n20} and Hutchings \cite{n21} and (b-c) UO$_{2.07}$ values. Figure 4a illustrates the excellent agreement between measured and reported data. ADPs for both uranium and oxygen in stoichiometric UO$_{2}$ gradually increase up to the melting point as a result of increasing thermal disorder. ADPs for the uranium site are lower than that of corresponding oxygen because of the lower magnitude of thermal vibration of the heavier uranium species compared to lighter oxygen.

\begin{figure}[!htb]
	\begin{center}
		\includegraphics[width=0.51\textwidth]{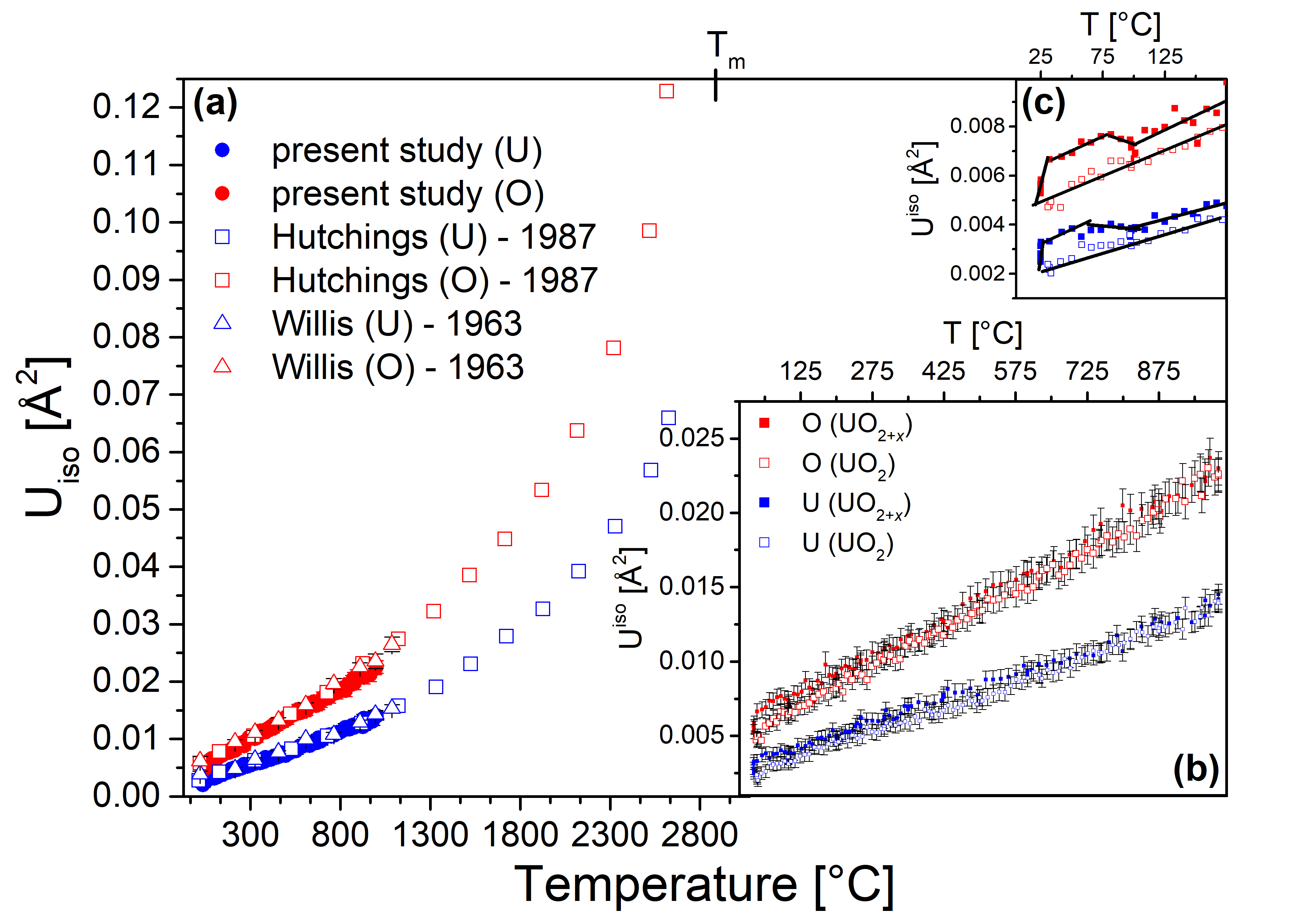}
		\caption{\label{Figure:4} (a) Evolution of the isotropic atomic displacement parameters (U$_{iso}$) of UO$_{2}$ from room temperature to the melting temperature (T$_{m}$) compared to reported findings from Willis \cite{n20} and Hutchings \cite{n21}. (b) Comparison of U$_{iso}$ values of UO$_{2}$ and UO$_{2.07}$ at various temperatures. (c) Low-temperature region from figure 4$b$ highlighting small changes in UO$_{2.07}$ U$_{iso}$ values compared to UO$_{2}$ values. Error bars are omitted in figure 4$c$ for clarity.}
	\end{center}
\end{figure}

A comparison of UO$_{2}$ and UO$_{2.07}$ ADPs reveals that oxidation causes an increase in ADPs, likely as a result of point defect accumulation. Both uranium and oxygen site ADPs of UO$_{2}$ and UO$_{2.07}$ are within experimental uncertainty for most of the temperature range studied. However, UO$_{2.07}$ values are consistently higher than UO$_{2}$ values at nearly all temperatures (Figure 4b). This indicates that UO$_{2.07}$ contains more static atomic disorder at any given temperature between 25--1000$\,^{\circ}\mathrm{C}$. A close inspection of the ADPs at low temperatures (Figure 4c) reveals that the difference between the UO$_{2}$ and UO$_{2.07}$ data exceeds the limits of experimental uncertainty between $\sim$25--100$\,^{\circ}\mathrm{C}$. ADPs of UO$_{2.07}$ in this region increase unusually fast, maintain higher values, and then decrease slightly at $\sim$100$\,^{\circ}\mathrm{C}$. This unique behavior is attributed to the hidden U$_{4}$O$_{9}$ phase and the UO$_{2+x (s)}$+$\alpha$-U$_{4}$O$_{9-y}$ $\to$ UO$_{2+x (s)}$+$\beta$-U$_{4}$O$_{9-y}$ phase boundary. The $\alpha$-to-$\beta$ phase transition temperature likely occurs at $\sim$80--90$\,^{\circ}\mathrm{C}$ based on Figure 4c. This value is in good agreement with the established phase diagram (Figure 1). No clear evidence for a UO$_{2+x (s)}$+$\beta$-U$_{4}$O$_{9-y}$ $\to$ UO$_{2+x (s)}$ transition was observed at higher temperatures from inspection of the ADP in Figure 4b because of the relatively large error bars.

In order to better probe the UO$_{2+x (s)}$+$\beta$-U$_{4}$O$_{9-y}$ $\to$ UO$_{2+x (s)}$ phase boundary, the evolution of the ADPs was compared to Debye models. Debye theory, strictly speaking, only applies to monatomic crystals. However, it has been successfully applied to well-behaved systems and simple ionic crystals in order to predict the temperature evolution of ADPs by estimating the ADP contribution from dynamic thermal displacements \cite{n22}. The application of Debye theory is beneficial because deviations in ADP trends from Debye predictions often indicate the presence of static atomic disorder that arises from defects or phase transformations. The temperature evolution of the ADPs was calculated using the expression:
\begin{displaymath}U_{T}=\frac{3 \hbar^{2}  T}{M k_{b}  \theta_{D}^{2}} \Big[\phi(\frac{\theta_{D}}{T})+\frac{1}{4} \frac{\theta_{D}}{T}\Big]+A
\end{displaymath}
where $U_{T}$ is the isotropic ADP at temperature $T$, $\hbar$ is the reduced Plank constant, $M$ is the mass of the vibrating atomic species, $k_{b}$ is the Boltzmann constant, $\theta_{D}$ is the Debye temperature of the material (125$\,^{\circ}\mathrm{C}$ for UO$_{2}$), $A$ is an arbitrary y-axis offset constant that accounts for intrinsic static atomic disorder in the system, and the function $\phi(x)$ is defined as:
\begin{displaymath}\phi(x) = \frac{1}{x}  \int_{0}^{x}\frac{x'}{e^{x^{'}}-1} dx^{'}\end{displaymath}

Willis \cite{n20} modeled the ADPs of stoichiometric UO$_{2}$ and concluded that the Debye model best represents the vibrations of heavy metal atoms and not lighter oxygen atoms. As such, data used for this modeling procedure were limited to uranium ADPs for UO$_{2}$ and UO$_{2.07}$. Moreover, only high-statistics data from the 30-minute measurements were used because they have much smaller error bars and enable more accurate assessment of small changes in ADP behavior. Figure 5 shows the comparison of the uranium ADP data to the Debye model predictions. The results show that Debye theory predicts the evolution of ADPs for UO$_{2}$ extremely accurately between the Debye temperature ($\sim$100$\,^{\circ}\mathrm{C}$) and the highest temperature (1000$\,^{\circ}\mathrm{C}$). This is in contrast to the UO$_{2.07}$ data, which show increasing deviation from the Debye prediction (red line in Figure 5) between $\sim$350 and 425$\,^{\circ}\mathrm{C}$. ADPs at higher temperatures ($>$350$\,^{\circ}\mathrm{C}$) are lower than the Debye prediction, which indicates that there is less static atomic disorder in the material at these elevated temperatures than what is predicted from fitting the lowest temperature data point (100$\,^{\circ}\mathrm{C}$).

\begin{figure}[!htb]
	\begin{center}
		\includegraphics[width=0.4\textwidth]{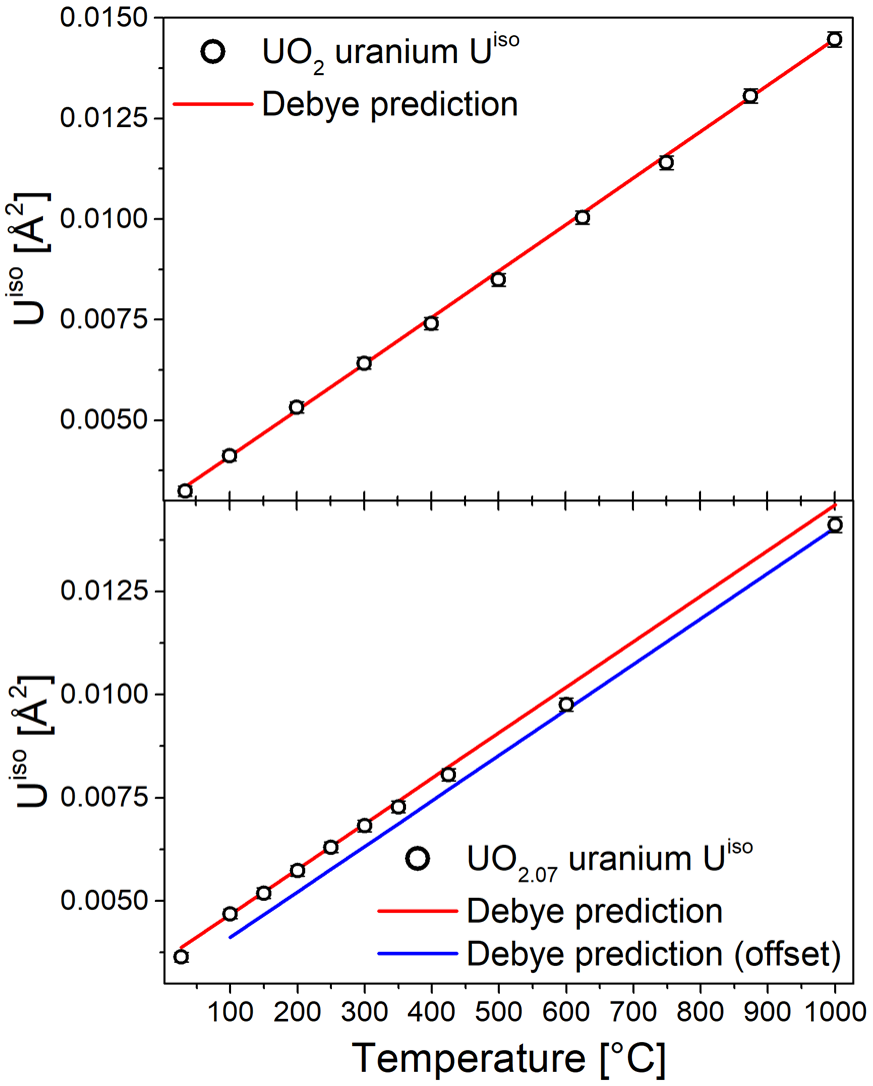}
		\caption{\label{Figure:5} Evolution of the uranium isotropic atomic displacement parameter (U$_{iso}$) for UO$_{2}$ and UO$_{2.07}$ compared with Debye model predictions. Red lines denote U$_{iso}$ trends fitted to the first few data points at or above the Debye temperature ($\sim$100$\,^{\circ}\mathrm{C}$). The blue line was fit to the highest temperature data point of UO$_{2.07}$ (1000$\,^{\circ}\mathrm{C}$) with an offset along the y-axis to account for a decrease in static atomic disorder.}
	\end{center}
\end{figure}

A second Debye model trendline with a negative y-axis offset (blue line in Figure 5) was fit to the ADP of UO$_{2.07}$ at 1000$\,^{\circ}\mathrm{C}$ in order to account for the decrease in static atomic disorder in the material. A comparison of the two trendlines and the ADP data for UO$_{2.07}$ suggests that the ADP data begin to deviate from the red line towards the blue line near $\sim$425$\,^{\circ}\mathrm{C}$. One explanation for this behavior is that the material crosses the UO$_{2+x (s)}$+$\beta$-U$_{4}$O$_{9-y}$ $\to$ UO$_{2+x (s)}$ phase boundary. The co-existence of UO$_{2+x (s)}$ and $\beta$-U$_{4}$O$_{9-y}$ phases increases microstrain at phase domain interfaces and increases static atomic disorder because the $\beta$-U$_{4}$O$_{9-y}$ phase contains cuboctahedral defect arrangements \cite{n13}. This explains why the initial Debye trendline is offset to higher ADP values: Upon transitioning to single-phase UO$_{2+x}$ (UO$_{2+x (s)}$) at higher temperatures, heterogeneous microstrain is mitigated. It is also possible that oxygen interstitials are more easily incorporated in UO$_{2+x (s)}$ because of thermal expansion effects. The easier incorporation of interstitials decreases static atomic disorder at temperatures greater than $\sim$600$\,^{\circ}\mathrm{C}$ and the Debye trendline is offset to lower ADP values. Further evidence for the UO$_{2+x (s)}$+$\beta$-U$_{4}$O$_{9-y}$ $\to$ UO$_{2+x (s)}$ transition was derived from analysis of short-to-intermediate range structures by PDF analysis.

\subsection{3.3. Pair Distribution Function (PDF) Analysis – Short-Range Structures}
Oxidation of UO$_{2}$ proceeds primarily through the incorporation of oxygen interstitials in the fluorite structure. Numerous experimental \cite{n23,n24,n25,n26,n27} and computational studies \cite{n1,n28,n29,n30,n31} have demonstrated that interstitials are not distributed at random, but rather form complex defect clusters. Inter- and intra-cluster atomic arrangements dictate phase stability in hyper-stoichiometric uranium oxides, especially U$_{4}$O$_{9}$ phases \cite{n13}. Therefore, short- and intermediate-range atomic arrangements were probed using neutron PDF analysis.
A PDF is an interatomic distance map of a material wherein each peak is made up of one or more atom-atom pair correlations. Figure 6 shows the PDFs of UO$_{2}$ and UO$_{2.07}$ at various temperatures. As denoted by the labels above the PDF peaks, the first four peaks largely comprise first nearest-neighbor (1-NN) U-O, O-O, and U-U correlations. The lone exception is the third peak, which is made up of overlapping 1-NN U-U and second nearest-neighbor O-O correlations. PDF peaks are relatively sharp and well defined at low temperatures but broaden significantly at higher temperatures. Broadening and concomitant lowering of peak intensities is induced primarily by thermal disorder. Besides general peak broadening behavior, Figure 6 shows that the local structures of UO$_{2}$ and UO$_{2.07}$ are extremely similar. The local atomic configurations of U$_{4}$O$_{9}$ phases are known to be distinct from that of fluorite-type UO$_{2+x (s)}$; therefore, U$_{4}$O$_{9}$ phases and their associated defect clusters, which co-exist with UO$_{2+x (s)}$ at low temperatures ($<$ 600$\,^{\circ}\mathrm{C}$) in UO$_{2.07}$, must contribute little to the overall PDF signal intensity. This is in agreement with the diffraction data, which showed no evidence of U$_{4}$O$_{9}$-type domains.

\begin{figure*}[!htb]
	\begin{center}
		\includegraphics[width=\textwidth]{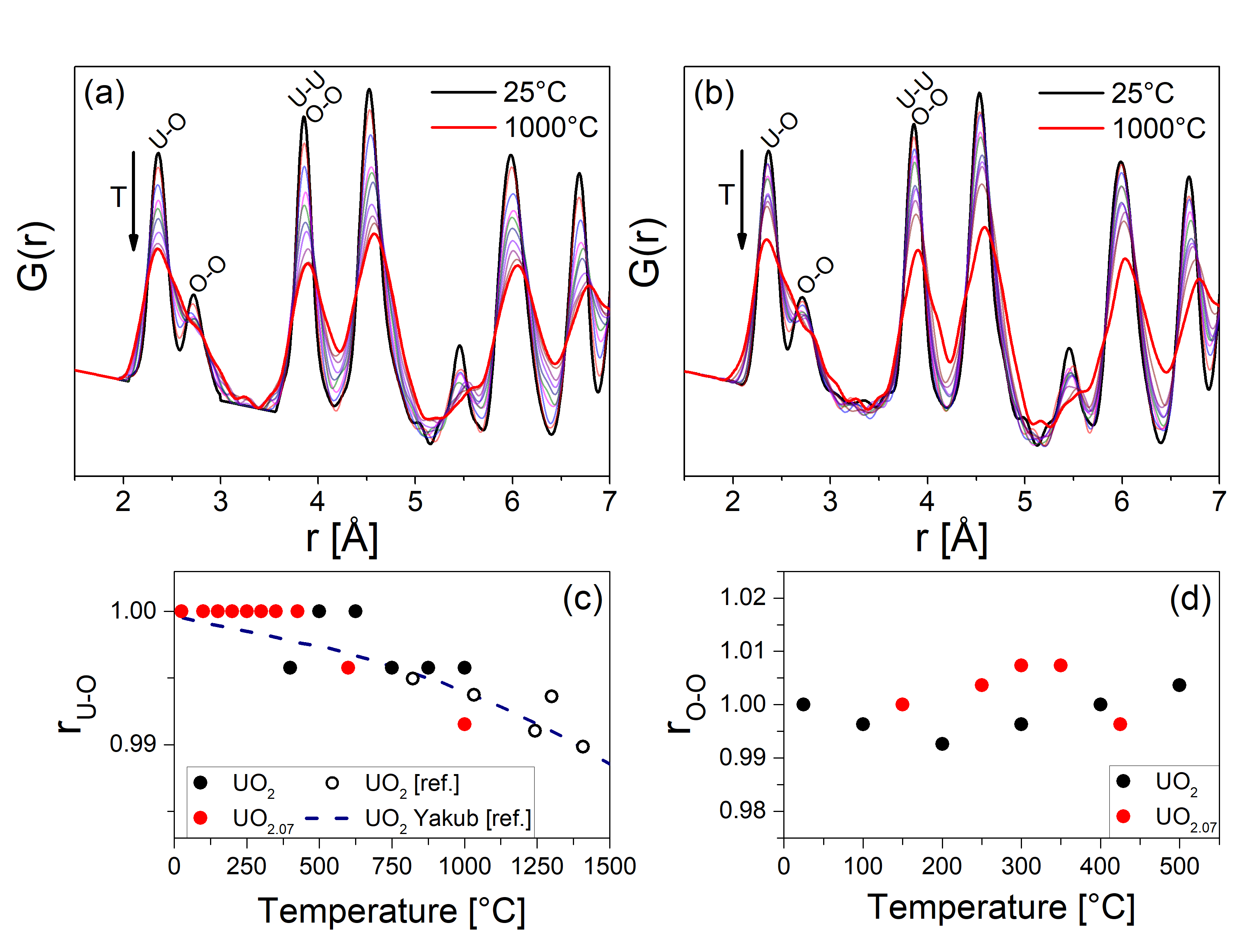}
		\caption{\label{Figure:6} Evolution of low-$r$ region of pair distribution functions (PDFs) of (a) UO$_{2}$ (b) UO$_{2.07}$ with increasing temperature. Labels are used to denote the first 4 peaks in the PDFs and the arrows illustrate how peak intensities decrease with increasing temperature. PDFs collected at temperatures 25--1000$\,^{\circ}\mathrm{C}$ are shown as colored curves with varying peak intensities. Peak maxima were used to plot the relative change in mean first nearest-neighbor (c) U-O and (d) O-O distances relative to room-temperature values. Black and red filled circles represent UO$_{2}$ and UO$_{2.07}$ data from the present study, respectively. Black open circles and the dashed line represent experimental data from Skinner \textit{et al.} \cite{n5} and molecular dynamics predictions referenced in \cite{n5}, respectively. Filled red circles and filled black circles overlap at 25$\,^{\circ}\mathrm{C}$, 100$\,^{\circ}\mathrm{C}$, 200$\,^{\circ}\mathrm{C}$, and 300$\,^{\circ}\mathrm{C}$ in (c), and at 25$\,^{\circ}\mathrm{C}$ and 100$\,^{\circ}\mathrm{C}$ in (d). O-O distances were not recorded above 500$\,^{\circ}\mathrm{C}$ because the first nearest-neighbor U-O and O-O peaks merge and O-O peak maxima were not clearly distinguishable at those temperatures.}
	\end{center}
\end{figure*}

When correlations are isolated, as is the case for 1-NN U-O and 1-NN O-O peaks, mean interatomic distances can be approximated by measuring the position of peak maxima in the PDF. Figures 6c-d show the evolution of 1-NN U-O and O-O distances for both UO$_{2}$ and UO$_{2.07}$ with varying temperature. Mean O-O values were limited to data collected at relatively low temperatures ($<$550$\,^{\circ}\mathrm{C}$) because the 1-NN O-O peak merged with the 1-NN U-O peak above $\sim$550$\,^{\circ}\mathrm{C}$ and it was not possible to accurately determine the position of the peak maximum at these temperatures. Results show that the 1-NN U-O distance ($\sim$2.36 \AA) for both UO$_{2}$ and UO$_{2.07}$ remains approximately fixed until $\sim$600$\,^{\circ}\mathrm{C}$ when the interatomic distance begins to decrease, despite the concurrent expansion in unit cell volume. Published results from molecular dynamics (MD) simulations \cite{n5} (dashed line in Figure 6) suggest a gradually decreasing U-O distance from $\sim$25$\,^{\circ}\mathrm{C}$ and approximate the experimental data quite well considering that there is scatter in data from both the present study and from the study of Skinner et al. \cite{n5}. Higher-temperature data from Skinner \textit{et al.} was derived from synchrotron X-ray PDF analysis (black open circles in Figure 6). Collectively, the results suggest that the Yakub MD potential is reliable for this application and the 1-NN U-O contraction that occurs at lower temperatures ($<$600$\,^{\circ}\mathrm{C}$) is too small to be experimentally detected.

The contraction of the 1-NN U-O distance with increasing temperature has been attributed to increasing disorder of UO$_{8}$ cubic polyhedra \cite{n5}. Thermal disorder and oxygen migration in the materials presumably promotes the co-existence of slightly over- and under-coordinated UO$_{8\pm x}$ polyhedra containing slightly shorter and longer U-O distances. It is possible that shorter U-O distances occur in larger concentrations and this results in an overall contraction of the mean 1-NN U-O distance. Whether these distorted configurations are stable or exist as transient states remains unknown owing to the inherent time-averaged nature of scattering measurements. Some researchers have shown that the very local structure of UO$_{2}$ can be accurately represented by $Pa\overline{3}$ local atomic arrangements at 1000$\,^{\circ}\mathrm{C}$ \cite{n3}. This type of local arrangement yields two distinct U-O distances, one shorter and one longer, rather than just one average U-O distance (as for $Fm\overline{3}m$) and results in the introduction of octahedral polyhedra into the structure. In order to agree with the long-range $Fm\overline{3}m$ symmetry of UO$_{2}$ from diffraction, it has been proposed that $Pa\overline{3}$ arrangements form as nano-domains that are modulated over longer length scales in order to yield an averaged fluorite-type arrangement \cite{n32}.

1-NN O-O distances ($\sim$2.72 \AA) for both UO$_{2}$ and UO$_{2.07}$ exhibit significant scatter over the entire temperature range. It is emphasized that the reported O-O distances merely represent peak maxima and likely do not correspond to true mean O-O distances. It is therefore possible that two or more characteristic O-O distances make up the 1-NN O-O peak and this causes the observed scatter. The non-linear changes in O-O distance can also result from the existence of a continuum of O-O distances that occur because the oxygen sublattice is very prone to disorder. In order to quantify the extent of atomic disorder and the very small differences between UO$_{2}$ and UO$_{2.07}$ short-range configurations, the PDFs were fit with simple fluorite-structure models using small-box PDF modeling.

PDFs of both UO$_{2}$ and UO$_{2.07}$ were well represented by ideal fluorite structures at all temperatures. As with Rietveld refinement, PDF fits of UO$_{2.07}$ were not improved upon adding interstitials ($x$ = 0.07) into the fluorite structure model at 4$\textit{b}$ octahedral sites. Figure 7 shows a comparison of PDF fit results for UO$_{2}$ and UO$_{2.07}$ data collected at room temperature. The figure shows that both PDFs are fit very well as illustrated by the green difference curves with low overall intensity. UO$_{2.07}$ fit results are in agreement with Rietveld refinement findings and show that structural features associated with U$_{4}$O$_{9}$-type phases are strongly suppressed at short length scales.

Attempts made to fit $\alpha$-type and $\beta$-type U$_{4}$O$_{9}$ phases based on proposed structural models \cite{n13} were initially unsuccessful. The two U$_{4}$O$_{9}$-type phases were each fit to UO$_{2.07}$ data as components of two-phase mixtures in conjunction with a simple fluorite-type UO$_{2.00}$ phase. However, U$_{4}$O$_{9}$ phases contain a much greater number of refineable parameters. This resulted in model instability when fitting two-phase mixtures because the relative intensity of the U$_{4}$O$_{9}$-phase was very low compared to that of the fluorite-type phase. Refinements performed with a very limited number of U$_{4}$O$_{9}$-phase refineable parameters yielded maximum U$_{4}$O$_{9}$ phase fractions of $\sim$15\% relative to the fluorite-type phase (85\%). These values are significantly smaller than phase fractions expected from phase diagram tie-line constructions for UO$_{2.07}$ at 25$\,^{\circ}\mathrm{C}$ and confirms that structural features associated with U$_{4}$O$_{9}$ phases are extremely weak.

\begin{figure}[!htb]
	\begin{center}
		\includegraphics[width=0.49\textwidth]{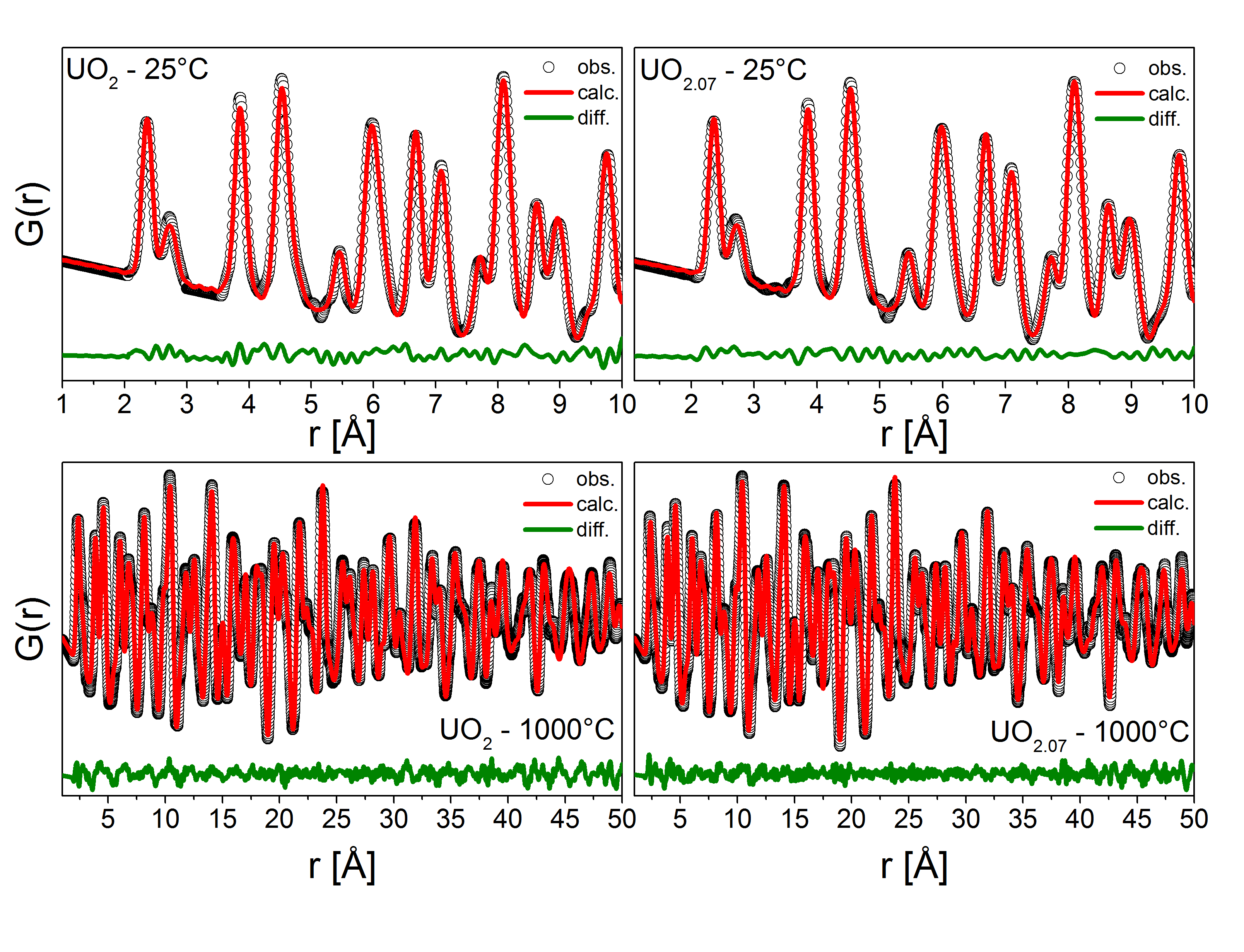}
		\caption{\label{Figure:7} Comparison of small-box refinement fits to the pair distribution functions (PDFs) of UO$_{2}$ and UO$_{2.07}$. The top frames show results from only fitting the very local structure (1--10\AA) at room temperature. The bottom frames illustrate results from fitting the entirety of the PDF range (1--50\AA) at 1000$\,^{\circ}\mathrm{C}$.}
	\end{center}
\end{figure}

Evidence for U$_{4}$O$_{9}$ features and phase boundaries was further probed by monitoring goodness-of-fit parameters, $R_{w}$, with varying temperature. For this analysis, both UO$_{2}$ and UO$_{2.07}$ PDFs were fit with a pristine fluorite structure model with no interstitials. Figure 8 shows the evolution of $R_{w}$ for UO$_{2}$ and UO$_{2.07}$ with increasing temperature. The fit to UO$_{2}$ data improves steadily with increasing temperature, likely because increased thermal disorder smoothens out sharp PDF features and facilitates peak fitting. Results for UO$_{2.07}$ show a similar progression except at 600 and 1000$\,^{\circ}\mathrm{C}$ where the trend deviates. Relative to the UO$_{2}$ trend, the $R_{w}$ values of UO$_{2.07}$ increase above 425$\,^{\circ}\mathrm{C}$, indicating a worsening fit. Comparison of the difference curves for UO$_{2}$ and UO$_{2.07}$ at 425, 600, and 1000$\,^{\circ}\mathrm{C}$ (Figure 7) did not illustrate a clear cause for the sharp deviation. To the naked eye, the UO$_{2}$ and UO$_{2.07}$ difference curves are very similar at 1000$\,^{\circ}\mathrm{C}$ despite the $R_{w}$ values being significantly different. The most likely cause for the deviation in the $R_{w}$ trend for UO$_{2.07}$ is the UO$_{2+x (s)}$+$\beta$-U$_{4}$O$_{9-y}$ $\to$ UO$_{2+x (s)}$ transition. The worsening fit at the highest temperatures indicates that the UO$_{2+x (s)}$ phase in the UO$_{2+x (s)}$+$\beta$-U$_{4}$O$_{9-y}$ mixture can be approximated as UO$_{2.00}$, in agreement with the phase diagram. At high-temperatures ($>$600$\,^{\circ}\mathrm{C}$), $\beta$-U$_{4}$O$_{9-y}$ domains dissolve and the UO$_{2.07}$ material becomes single-phase with an oxygen interstitial ordering scheme that is distinct from that of U$_{4}$O$_{9}$ phase.

Despite the identification of the phase boundary, specific structural changes are not clearly distinguishable from the PDFs, and differences between data and the fitted models do not exhibit any strong $r$-dependence. In other words, changes occurring in the PDF upon transitioning from mixed UO$_{2+x (s)}$+$\beta$-U$_{4}$O$_{9-y}$ to single-phase UO$_{2+x (s)}$ are not specific to short- or intermediate-range structures. Atomic structure modifications likely occur in small amounts throughout the $r$ range. This weak $r$-dependence is expected when point defects exist in clustered arrangements, as is the case for hyper-stoichiometric UO$_{2+x}$. If point defects were instead accommodated randomly, the material would eventually tend towards an amorphous state (\textit{i.e.}, loss of long-range order) and changes in the PDF would be concentrated at higher $r$ values, which is not the case. A close comparison of fitted peak intensities in Figure 7 suggests that the largest differences between the three UO$_{2+x}$ phases (mixed-phase UO$_{2+x (s)}$+$\alpha$-U$_{4}$O$_{9-y}$, mixed-phase UO$_{2+x (s)}$+$\beta$-U$_{4}$O$_{9-y}$, and single-phase UO$_{2+x (s)}$) are relative changes in PDF peak intensities. This suggests that these distinct mixtures and phases are closely related and only differ by small changes in oxygen defect accommodation scheme \cite{n13}.

\begin{figure}[!htb]
	\begin{center}
		\includegraphics[width=0.35\textwidth,trim=2 2 2 2,clip]{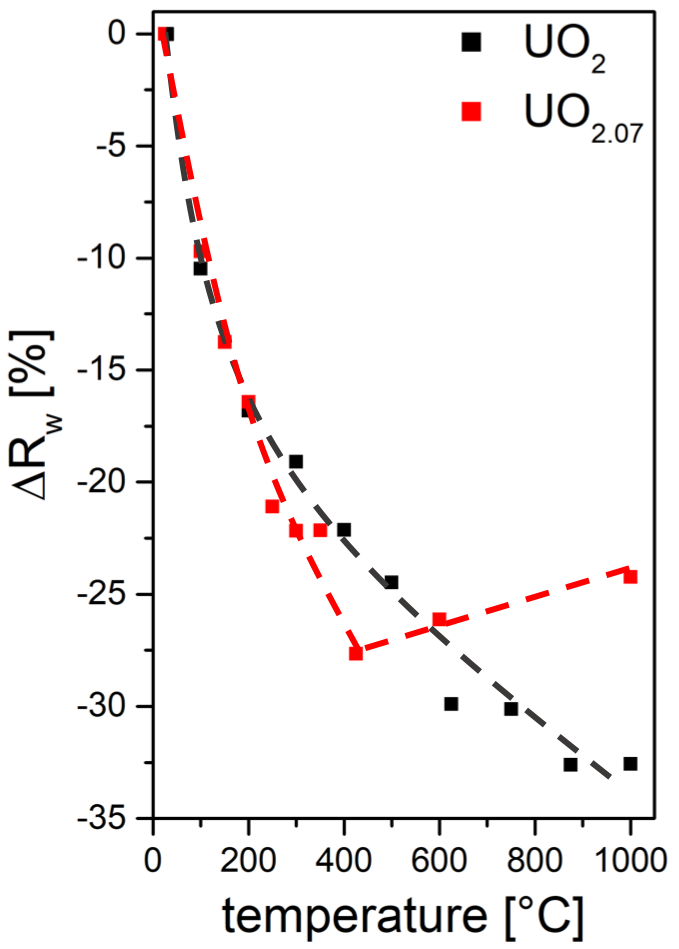}
		\caption{\label{Figure:8} Relative change of pair distribution function goodness-of-fit parameters, $\Delta$$R_{w}$, of UO$_{2}$ and UO$_{2.07}$ with increasing temperature. The increase in $\Delta$$R_{w}$ for UO$_{2.07}$ at the highest temperatures indicates that the fit worsens at 600$\,^{\circ}\mathrm{C}$ and 1000$\,^{\circ}\mathrm{C}$ relative to the fit at 425$\,^{\circ}\mathrm{C}$.}
	\end{center}
\end{figure}

Defect accommodation schemes in U$_{4}$O$_{9}$-type phases were elucidated previously by neutron PDF analysis \cite{n13}, but accurate assessment of defect clusters in single-phase UO$_{2+x}$ remains challenging owing to the diffuse nature of interstitials in UO$_{2+x}$ in the low O:M regime. As a test, the very local structures (1 \AA $<$ $r$ $<$ 10 \AA) of the PDFs of single-phase UO$_{2.07}$ were fit with established defect models from the literature. UO$_{2.07}$ defect models were approximated using 2$^{3}$ UO$_{2}$ supercells containing 2 additional oxygen atoms (overall composition of U$_{32}$O$_{66}$ or $\sim$UO$_{2.063}$). The two interstitials were arranged either at 4\textit{b} sites (octahedral model), as split di-interstitial defect clusters (di-interstitial model) \cite{n1}, or as 2:2:2-type defect clusters \cite{n25} (Willis model). Fitting these simple models to the PDF of UO$_{2.07}$ at 1000$\,^{\circ}\mathrm{C}$ yielded $R_{w}$ values of 0.086, 0.082, and 0.075 for octahedral, di-interstitial, and Willis models, respectively. Fitting to UO$_{2.07}$ data at 600$\,^{\circ}\mathrm{C}$ yielded $R_{w}$ values of 0.072, 0.068, and 0.064, respectively. These results indicate that the local structure of single-phase UO$_{2.07}$ at 600 and 1000$\,^{\circ}\mathrm{C}$ is most consistent with the 2:2:2 Willis defect cluster model; however, these results stem from fitting only the very local structure of the material (1 \AA $<$ $r$ $<$ 10 \AA) and do not consider the changes in intermediate- and long-range structures or changes in defect cluster geometry. In-depth modeling and analysis of point defect clusters was considered beyond the scope of the current study. Recent findings from Ma \textit{et al.} \cite{n33} demonstrate that it is very challenging to differentiate between different defect models when interpreting subtle changes in PDF peak intensities. Development of novel data modeling methods is needed in order to reconcile the remaining discrepancy in atomic ordering schemes between short- and long-range length scales.

\section{4. Conclusions}
UO$_{2}$ and UO$_{2.07}$ were characterized from 25--1000$\,^{\circ}\mathrm{C}$ using neutron total scattering in order to evaluate effects of temperature and phase boundaries on structural arrangements. Analyses of unit cell parameters showed that both materials exhibit very similar thermal expansion behavior and thermal expansion data lie along the upper bound of uncertainty for currently-accepted empirical models. Atomic displacement parameters of UO$_{2.07}$ show evidence for U$_{4}$O$_{9}$ phase boundaries in accordance with the established phase diagram despite the suppression of U$_{4}$O$_{9}$ superlattice peaks in the measured diffraction patterns. Pair distribution function (PDF) analysis revealed that the differences in local structure between UO$_{2}$ and UO$_{2.07}$ are very small and defect signatures in the PDFs are heavily obscured by thermal effects. 

\section{Acknowledgements}
This research was supported by the Office of Basic Energy Sciences of the U.S. Department of Energy as part of the Materials Science of Actinides Energy Frontier Research Center (DE-SC0001089). The research at ORNL's Spallation Neutron Source was sponsored by the Scientific User Facilities Division, Office of Basic Energy Sciences, US Department of Energy. JL acknowledges the financial support from the U.S. Department of Energy, Office of Nuclear Energy under a Nuclear Energy University Program (DE-NE0008440). R.I.P. acknowledges support from the U.S. Department of Energy (DOE) National Nuclear Security Administration (NNSA) through the Carnegie DOE Alliance Center (CDAC) under grant number DE-NA-0002006.

\bibliographystyle{apsrev4-1}
\bibliography{UO2_oxygen_defects.bib}

\end{document}

%% file: author.tex
\affiliation{Department of Nuclear Engineering, University of Tennessee, Knoxville, Tennessee, 37996}
\affiliation{Department of Civil and Environmental Engineering and Earth Sciences, University of Notre Dame, Notre Dame, Indiana 46556}
\affiliation{Department of Mechanical, Aerospace, and Nuclear Engineering, Rensselaer Polytechnic Institute, Troy, New York}
\affiliation{Chemical and Engineering Materials Division, Oak Ridge National Laboratory, Oak Ridge, Tennessee, 37831}
\author{Raul I.~Palomares\footnote{Corresponding Author: raul.i.palomares@gmail.com}} \affiliation{Department of Nuclear Engineering, University of Tennessee, Knoxville, Tennessee, 37996}
\author{Jennifer E.~Szymanowski} \affiliation{Department of Civil and Environmental Engineering and Earth Sciences, University of Notre Dame, Notre Dame, Indiana 46556}
\author{Tiankai~Yao} \affiliation{Department of Mechanical, Aerospace, and Nuclear Engineering, Rensselaer Polytechnic Institute, Troy, New York}
\author{Joerg~Neuefeind} \affiliation{Neutron Scattering Division, Oak Ridge National Laboratory, Oak Ridge, Tennessee, 37831}
\author{Ginger E.~Sigmon} \affiliation{Department of Civil and Environmental Engineering and Earth Sciences, University of Notre Dame, Notre Dame, Indiana 46556}
\author{Jie~Lian} \affiliation{Department of Mechanical, Aerospace, and Nuclear Engineering, Rensselaer Polytechnic Institute, Troy, New York}
\author{Maik~Lang} \affiliation{Department of Nuclear Engineering, University of Tennessee, Knoxville, Tennessee, 37996}
%
%
\vskip 0.25cm